\documentclass[12pt]{article}
\usepackage[margin=1in]{geometry}

\usepackage{amssymb,amsmath,amsthm,longtable}
\usepackage{bm,color,enumerate,framed,setspace,tikz,dsfont}
\usepackage[colorlinks,citecolor=blue]{hyperref}
\usepackage[round]{natbib}
\usepackage[nolists]{endfloat}
\usepackage[pagewise]{lineno}
\usepackage[algonl]{algorithm2e}

\newcommand*\patchAmsMathEnvironmentForLineno[1]{%
  \expandafter\let\csname old#1\expandafter\endcsname\csname #1\endcsname
  \expandafter\let\csname oldend#1\expandafter\endcsname\csname end#1\endcsname
  \renewenvironment{#1}%
     {\linenomath\csname old#1\endcsname}%
     {\csname oldend#1\endcsname\endlinenomath}}% 
\newcommand*\patchBothAmsMathEnvironmentsForLineno[1]{%
  \patchAmsMathEnvironmentForLineno{#1}%
  \patchAmsMathEnvironmentForLineno{#1*}}%
\AtBeginDocument{%
\patchBothAmsMathEnvironmentsForLineno{equation}%
\patchBothAmsMathEnvironmentsForLineno{align}%
\patchBothAmsMathEnvironmentsForLineno{flalign}%
\patchBothAmsMathEnvironmentsForLineno{alignat}%
\patchBothAmsMathEnvironmentsForLineno{gather}%
\patchBothAmsMathEnvironmentsForLineno{multline}%
}

\newcommand{\Mta}{$M_{t\alpha}$}
\newcommand{\ind}{\mathds{1}}
\renewcommand{\paragraph}[1]{}
\newcommand{\MCMC}{MCMC~}
\newcommand{\MCMCfirst}{Markov chain Monte Carlo (MCMC)~}
\newcommand{\LMM}{LMM~}%{latent multinomial model~}
\newcommand{\LMMfirst}{latent multinomial model (LMM)~}
\newcommand{\MHfirst}{Metropolis Hastings (MH)~}
\newcommand{\BRE}{BRE~}
\newcommand{\BREfirst}{band-read error (BRE)~}
\newcommand{\CJSfirst}{Cormack-Jolly-Seber (CJS)~}
\newcommand{\CJS}{CJS}
\newcommand{\CJSBRE}{CJS/BRE~}

\newcommand{\figurebox}[3][dummy]{}

\newtheorem{theorem}{Theorem}

\title{Extending the Latent Multinomial Model with Complex Error Processes and Dynamic Markov Bases}
\author{Simon J Bonner, Matthew R Schofield, Patrik Noren, and Steven J Price}

\begin{document}
\maketitle

% Start line numbering
%\linenumbers

% Start double spacing
%\doublespacing

\begin{abstract}
The \LMMfirst model 
%(\LMM)  %% Removed for Biometrika
of \citet{Link2010} provided a general framework for modelling mark-recapture data with potential errors in identification. Key to this approach was a \MCMCfirst 
%(\MCMC) %% Removed for Biometrika
scheme for sampling possible configurations of the counts true capture histories that could have generated the observed data. This \MCMC algorithm used vectors from a basis for the kernel of the linear map between the true and observed counts to move between the possible configurations of the true data. \citet{schofield2014} showed that a strict basis was sufficient for some models of the errors, including the model presented by \citet{Link2010}, but a larger set called a Markov basis may be required for more complex models. We address two further challenges with this approach: 1) that models with more complex error mechanisms do not fit easily within the \LMM and 2) that the Markov basis can be difficult or impossible to compute for even moderate sized studies. We address these issues by extending the \LMM to separately model the capture/demographic process and the error process and by developing a new \MCMC sampling scheme using dynamic Markov bases. Our work is motivated by a study of Queen snakes ({\it Regina septemvittata}) in Kentucky, USA, and we use simulation to compare the use of PIT tags, with perfect identification, and brands, which are prone to error, when estimating survival rates.  

%%% Local Variables: 
%%% mode: latex
%%% TeX-master: "../bre_manuscript_1"
%%% End: 

\end{abstract}


\section{Introduction}

\paragraph{Opening}
\label{sec:opening}

Standard models for data from studies of marked individuals require that researchers are able to identify marked individuals uniquely and without error. However, these assumptions may be violated in many ways. Researchers may misread marks and provide partial identifications based on visual sightings or poor quality photographs \citep{mcclintock2014,Morrison2011}, allelic dropout may lead to incorrect identifications from DNA samples \citep{Lukacs2005,Wright2009,Yoshizaki2011}, man-made tags may be lost or degrade \citep{cowen2006}, and natural marks may evolve over time \citep{Yoshizaki2012}. \citet{Link2010} described a new framework to allow for misidentification in mark-recapture data or, in their words, data that present a ``mangled and incomplete summary'' of the true capture histories. The framework depends on writing the observed capture history frequencies as a linear function of the true frequencies, described through the equation $\bm x=\bm A \bm n$, and was called the \LMMfirst \citep{Link2010}. Key to fitting this model was a novel \MCMCfirst sampling algorithm that used a subset of elements in the kernel of $\bm A$, $\ker(\bm A)$, to generate proposals for $\bm n$ in a Metropolis-Hastings update step. In particular,  \citet{Link2010} draw elements from a basis for $\ker(\bm A)$. 

\citet{schofield2014} showed that more care may be needed to ensure that the resulting Markov chains are irreducible and cover the entire sample space. \citet{Link2010} focused on one model of errors in closed populations, \Mta, and provided details of the specific basis used in the algorithm only for $T=2$ capture occasions. However, the discussion implied that the algorithm could be implemented with any basis of $\ker(\bm A)$ and could be applied whenever the observed counts can be written as a linear function of latent counts. \citet{schofield2014} proved that the specific basis chosen for \Mta~when $T=2$ does produce irreducible chains, but they showed that this is not true for all bases. They also provided examples of models with more complex error for which no set of linearly independent elements $\ker(\bm A)$ can produce irreducible chains. To address these issues they first provided a method for constructing a basis that is guaranteed to produce irreducible chains when the errors form simple corruptions (i.e., they corrupt the capture history for a single individual). This includes model \Mta. Furthermore, they showed that algorithms for more complex models can be constructed by adding elements to the basis to construct a so called Markov basis which guarantees that the resulting Markov chains are irreducible \citep{Diaconis1998}.

Although the use of Markov bases seems to address the problems with more complex models, two challenges remain. First, it can be difficult to describe the distribution of the latent frequencies when errors affect multiple individuals. This makes it difficult to cast the model in the \LMM framework. Second, the approach of computing Markov bases directly is impractical for realistic mark-recapture data sets. General methods for computing Markov bases have been developed and implemented in software packages like \texttt{4ti2}. However, the Markov bases for these models are so large or complex that they cannot be computed with current hardware. For the specific model we present in Section \ref{sec:data}, \texttt{4ti2} exceeded the memory on a computer with 8~GB of RAM when the experiment contained 5 capture occasions or more. 

We first develop an extension of the \LMM that separates the models of the population demographics/capture process and error process by introducing a second set of latent counts (Section \ref{sec:methods-1}), and then describe a new \MCMC algorithm using dynamic Markov bases to avoid explicit computation of the Markov bases (Section \ref{sec:methods-2}). Our work is motivated by a study of queen snakes ({\it Regina septemvittata}) in Jessamine County, Kentucky. Since 2013, snakes have been marked with subdermal passive integrated transponder (PIT) tags, and this method has several advantages. Identifications from PIT tags are almost 100\% reliable and PIT tags can be detected from a distance (up to 42~cm) so that snakes can be identified without physical capture and displacement of habitat \citep[see][]{connette2012}. The use of PIT tags may also increase the detection rates, but PIT tags are expensive costing \$5 per tag plus almost \$3000 per receiver. As an alternative, snakes may be marked with unique brands applied with handheld medical cautery units \citep{Winne2006}. The cost of brand snakes is negligible, but branded snakes must be physically captured to be identified and brands are easy to misread. The extensions of the \LMM we provide would allow us to model brand data while accounting for potential errors. We use simulations based on the available PIT tag and brand data to assess the impacts that these errors would have on the estimation of survival rates and discuss how these extensions can be applied to a broader range of models.

\section{The Latent Multinomial Model}

The \LMM of \citet{Link2010} accounts for possible errors in the data by recasting the mark-recapture model. Suppose that $I$ different capture histories could be observed during the study. If the population is homogeneous then the $I$-vector of counts, $\bm n$, recording the number of times each history was observed is a sufficient statistic. Unfortunately, the distribution of $\bm n$ depends on both the mark-recapture and error processes and may be difficult to compute directly. To make the likelihood tractable, the \LMM introduces a set of $J>I$ latent histories which identify the true captures for each individual and describe what errors occurred. Let $\bm x$ be the unobserved $J$-vector of counts for these latent histories. The likelihood can be defined by summing probabilities over all values of $\bm x$ consistent with $\bm n$. In particular, the \LMM assumes that there is a linear relationship between $\bm n$ and $\bm x$ so that $\bm n=\bm A \bm x$ for some known $I \times J$ matrix $\bm A$. The likelihood can then be computed as
\begin{equation}
\label{eqn:lhd-1}
\pi(\bm n|\bm \theta)
=\sum_{\bm x \in \mathbb N^J} \ind(\bm n=\bm A \bm x)\pi(\bm x|\bm \theta)
=\sum_{\bm x \in \mathcal F_{\bm n}}\pi(\bm x|\bm \theta)
\end{equation}
where $\ind(\cdot)$ is the indicator function and $\bm \theta$ the vector of parameters, $\mathcal F_{\bm n}=\{\bm x \in \mathbb N^J: \bm n=\bm A\bm x\}$ is the inverse image of $\bm n$ (called the $\bm n$-fibre in algebraic statistics), and $\mathbb N$ is the set of natural numbers including 0.

% For example, the latent history $01102$ might be used to denote that an individual was captured on occasions 2,3, and 5 but misidentified on occasion 5, as described above. The latent histories generally contain more events than the simple binary capture indicators so that there are more latent histories then observable histories, $J>I$, and there is a many-to-one map from the sample space of $\bm x$ to the sample space of $\bm n$. The distribution of $\bm n$ can then be computed by summing over all values of $\bm x$ in its inverse image 

As an example,  \citet{Link2010} considered an extension of the time-dependent, closed population model $M_t$ of \citet{Otis1978a} called \Mta{}. This model was first described by \citet{Yoshizaki2011} and makes two assumptions regarding errors: 1) that captured individuals are correctly identified with probability $\alpha$ independent of all other events and 2) that the identities resulting from errors are unique and do not match the marks of other individuals in the population or the identities generated by previous errors. The second assumption implies that each error removes a single observation from one individual's true capture history and produces a new observed history containing a single capture event.

Although equation (\ref{eqn:lhd-1}) may make it easier to compute the likelihood function in theory, $\mathcal F_{\bm n}$ is often so large that exact computation is not practical. Instead, \citet{Link2010} proposed to sample from the joint posterior distribution of both the latent vector, $\bm x$, and the model parameters, $\bm \theta$. 
% The sum in equation (\ref{eqn:lhd-1}) is then estimated by marginalizing the posterior distribution with respect to $\bm x$. 
The specific \MCMC algorithm uses a block \MHfirst approach and the primary challenge lies in constructing proposals of $\bm x|\bm \theta$ which are likely to lie inside the fibre. The algorithm starts by defining a lattice basis for the kernel of $\bm A$; that is, a set  $\mathcal B=\{b_1,\ldots,b_K\}$ such that any $\bm b$ in $\ker(\bm A)$ with integer entries, $\bm b \in \ker(\bm A) \bigcap \mathbb Z^J$, can be written as a linear combination of the elements of $\mathcal B$ with integer coefficients $c_1,\ldots,c_K \in \mathbb Z$. Integer multiples of the elements from the lattice basis are then added to the current value of $\bm x$ one-at-a-time to generate new proposals that are accepted or rejected before continuing to the next element. The magic of this approach is that  $\bm A(\bm x + c\bm b)=\bm A \bm x$ for any $\bm x \in \mathcal F_{\bm n}$, $\bm b \in \ker(\bm A)$, and $c \in \mathbb Z$ so that the proposal $\bm x^{(\mbox{prop})}=\bm x + c_k\bm b_k$ also has integer entries and is guaranteed to satisfy the linear constraint. Note, however, that $\bm x^{(\mbox{prop})}$ may still fall outside $\mathcal F_{\bm n}$ since there is no guarantee that its entries will all be non-negative.

\citet{Link2010} implied that Markov chains constructed with this algorithm would connect all elements in $\mathcal F_{\bm n}$ and hence be irreducible. \citet{schofield2014} showed that this is true for model $M_{t\alpha}$ provided that the right lattice basis is chosen and extended this result to a broader class of models in which errors constitute simple corruptions. However, they also provided examples of more complicated models for which the algorithm does not produce irreducible Markov chains. The central problem is that some pairs of elements in $\mathcal F_{\bm n}$ may be connected by the algorithm above only by passing through values of $\bm x$ containing negative entries. 
% That is, for some $\bm x_1,\bm x_2 \in \mathcal F_{\bm n}$ and any sequence of elements $\bm b_1,\ldots,\bm b_K \in \mathcal B$ and $c_1,\ldots,c_K \in \mathbb Z$ such that 
% \[
% \bm x_1=\bm x_2 + \sum_{k=1}^K c_k \bm b_k
% \]
% the partial sum
% \[
% \bm x_2 + \sum_{k=1}^{K'} c_k \bm b_k
% \]
% contains a negative entry and lies outside $\mathcal F_{\bm n}$ for some $K'<K$ . 
Since these elements lie outside of $\mathcal F_{\bm n}$ and have zero probability under the posterior, the chain will never be able to move between $\bm x_1$ and $\bm x_2$ and will not be irreducible.

Irreducible chains can always be produced by adding linear combinations of all elements in $\mathcal B$ simultaneously instead of adding elements one at a time, but the resulting proposals are likely to contain negative entries and \citet{Diaconis1998} reported that this method is not efficient. Instead, \citet{Diaconis1998} suggested using the one-at-a-time algorithm but drawing the elements from a larger subset $\mathcal M \subset \ker(\bm A)$ chosen to ensure that it is possible to move between any two elements of $\mathcal F_{\bm n}$. \citet{Diaconis1998} called $\mathcal M$ a Markov basis and the elements of $\mathcal M$ moves. We consider the special case of the algorithm presented by \citet{schofield2014} in which one element is selected from the Markov basis on each iteration of the \MCMC algorithm and either added to or subtracted from the current configuration without a multiplier. Details are given in Algorithm \ref{alg:1}.

\paragraph{Objectives}
\DontPrintSemicolon

\begin{algorithm}
  \caption{\MCMC algorithm for sampling from the joint posterior distribution of $\bm \theta$ and $\bm x$ given a fixed Markov basis, $\mathcal M$.}
  Define a Markov basis, $\mathcal M$.\;
Initialize $\bm \theta_1^{(0)}$, $\bm \theta_2^{(0)}$, and $\bm x^{(0)}$ so that $\bm n = \bm A \bm x^{(0)}$.\;

\BlankLine

Set $k=1$.\;

%\BlankLine

%\Repeat{convergence}{
1) Update $\bm \theta$ conditional on $\bm x^{(k-1)}$. Call the result $\bm \theta^{(k)}$.\;
%\BlankLine
2) Update $\bm x$ conditional on $\bm \theta^{(k)}$.\;
\Indp 
%\Indp
a) Select $\bm b \in \mathcal M$ and $c \in \{-1,1\}$\; %\mathcal C(\bm b)$.\;
b) Set $\bm x^{\mathrm{prop}}= \bm x^{(k-1)} + c\bm b$.\;
c) Calculate the Metropolis acceptance probability: 
\[
r = \min\left\{1,\frac{\pi(\bm n,\bm x^{\mathrm{prop}}|\bm \theta^{(k)})}
  {\pi(\bm n,\bm x^{(k-1)}|\bm \theta^{(k)})}\cdot 
  \frac{q(\bm x^{(k-1)}|\bm x^{\mathrm{prop}})}
  {q(\bm x^{\mathrm{prop}}|\bm x^{(k-1)})}
\right\}
\]
where $q(\bm x'|\bm x)$ is the probability of proposing $\bm x'$ given the current state $\bm x$.\;
d) Set $\bm x^{(k)}=\bm x^{\mathrm{prop}}$ with probability $r$. Otherwise, set $\bm x^{(k)} = \bm x^{(k-1)}$.\;
%\Indm
\Indm
\BlankLine
3) Increment $k$.\;
\BlankLine
%}

%%% Local Variables: 
%%% mode: latex
%%% TeX-master: "../bre_manuscript_1"
%%% End: 

  \label{alg:1}
\end{algorithm}

%%% Local Variables: 
%%% mode: latex
%%% TeX-master: "../bre_manuscript_1"
%%% End: 

\section{Data and Models}

\label{sec:data}

\paragraph{Data}
The data we consider comes from a study of queen snakes conducted along Little Hickman Creek in Jessamine County, Kentucky. An initial sample of 61 snakes was captured and marked in the fall of 2013 and a second sample of 41 snakes was marked  in the spring of 2014. All snakes were implanted with PIT tags and a subset of 73 snakes were also branded with unique marks as described in \citet{Winne2006}. In the summer of 2014, two technicians visited the site to locate and identify snakes approximately every two weeks. On each visit the technicians conducted searches using a PIT receiver and attempted to physically capture any snakes that were detected so that their brands could be read. The 102 snakes were re-encountered 191 times in total, an average of 1.87 per snake. The researchers conducting the study are primarily interested in modelling the survival and movements of the snakes in this population and in understanding the individual and population level impacts of snake fungal disease, am emerging pathogen about which little is currently known \citep{alender2013,sleeman2013}. For illustration, we focus on modelling the snakes' apparent over-wintering survival, the probability that a snake marked in the fall of 2013 is still in the population in 2014. 

\paragraph{Model}

Previous studies have found that snakes may expel PIT tags \citep[e.g.][]{roark2000} and some loose tags were found at the study site. However, we believe that the rate of expulsion is small and there is no reason to think that PIT tags are ever misidentified.  With these assumptions capture histories formed using the PIT tag encounters can be modelled with standard Cormack-Jolly-Seber type models ignoring potential identification errors or tag loss \citep[see][and references therein]{lebreton1992,seber2002a,Williams2002}. On the other hand, the brands can be difficult to read and the identification of physically captured snakes is prone to error. A total of 9 branded snakes were recaptured physically during the summer of 2014. By comparing with the PIT tag records we knew that the technician who had originally branded the snakes  identified 8 of 9 (89\%) correctly while the second technician identified only 6 of 9 (67\%) correctly. The small number of physical recaptures did not allow us to compare results based on the PIT tag and brand data directly. Instead, we examine the feasibility of branding snakes by analyzing simulated brand data generated with error rates matching those observed from the two technicians. 
% We then compare inferences from the analysis of the original PIT tag data and the simulated data both with standard CJS models and the new methods we develop to account for potential errors.

The specific model we consider both for generating and analyzing the simulated brand data combines the standard \CJSfirst model for the capture process and the \BREfirst model defined \citet{schofield2014}. Suppose that researchers visit a location on $T$ occasions. On each visit they capture a number of unmarked individuals, mark them, and return them to the population. At the same time, the researchers also conduct visual surveys  to identify previously marked individuals. The assumptions of the \BRE model are: 
\begin{enumerate}
\item Resighted individuals are correctly identified with probability $\alpha$ on each occasion,
\item Errors cause one marked individual to be misidentified as another marked individual.
\label{ass:bre2}
\item Each individual can only be involved in one event on any one occasion. In particular, it is not possible to resight individual $i$ and to mistake another individual for individual $i$ on the same occasion.
\end{enumerate}
Assumption \ref{ass:bre2} contrasts directly with the assumptions of model \Mta{} and is justified by the differences between man-made marks and natural marks. Model \Mta{} is intended for use with natural marks including genotypes and pigmentation patterns. The set of possible natural marks is usually unknown and the number of possible marks is so large that it unlikely for an error to reproduce the identity of another individual exactly. On the other hand, the \BRE model is intended for use with man-made marks. The set of possible marks is known when using man-made marks, and this means that erroneous sightings of marks which have never been released can be detected and removed from the data prior to the analysis. The only errors that cannot be detected occur when one marked individual is mistaken for another marked individual. The third assumption simplifies the model and we plan to relax this in future work. Details on the likelihood for this model are provided in Section \ref{sec:methods-1} after we introduce the new modelling framework. Note that erroneous sightings of marks that were released is only appropriate when estimating survival, as we do here, and leads to biased estimates of abundance \citep[see][]{white2001,mcclintock2014}. 

%%% Local Variables: 
%%% mode: latex
%%% TeX-master: "../bre_manuscript_1"
%%% End: 

%\section{Methods}

\section{Extended Framework}

\label{sec:methods-1}

\paragraph{Introduction}

The framework of \citet{Link2010} focused on models in which $\bm x$ follows a multinomial distribution. This is the case for model \Mta{} and although they suggested that the methods could be applied more generally examples were not provided. The \CJSBRE model does not result in a multinomial distribution for $\bm x$, and it is difficult to determine the density of $\bm x$ for the \CJSBRE model explicitly, making it hard to apply the \LMM directly.

% \begin{itemize}
% \item \citet{Link2010} focuses on models in which $\bm x$ follows a multinomial distribution
% \item this is the case for the example model they consider, \Mta{}
% \item he does suggest that the same methods may be applied when $\bm x$ does not follow a multinomial distribution, but does not follow this further
% \item in practice, it may be difficult to construct the density of $\bm x$ explicitly for more complicated forms of errors
% \end{itemize}

\paragraph{New Formulation}

To address this, we extend the \LMM to include a second vector of latent counts which allows the mark-recapture process and the error mechanism to be modelled separately. Suppose, for example, that an experiment has $T=2$ occasions and individual $i$ is captured on both occasions, correctly identified on the first occasion, and identified as an entirely new individual on the third occasion (this is the error mechanism for model \Mta{}). In the terminology of \citet{Link2010}, individual $i$ would have latent history $\bm \nu_i=12$ and would produce the recorded histories $\bm \omega_{i1}=10$ and $\bm \omega_{i2}=01$. The original \LMM assigns probabilities to the latent histories, $\bm \nu_i$, directly by simultaneously modelling the capture and error processes. Our formulation introduces a second latent history, $\bm \xi_i$, identifying the occasions on which the individual was truly captured but ignoring the errors. The new latent history would be $\bm \xi=11$ since the individual was truly captured on both occasions. We then model the joint distribution of $\bm \nu_i$ and $\bm \xi_i$ by assigning probabilities first to $\bm \xi_i$ and second to $\bm \nu_i$ given $\bm \xi$. We distinguish between the two sets of latent histories by calling $\bm \nu_i$ the latent error history and $\bm \xi_i$ the latent capture history. 

% Generally, we assume that we have three sets of capture histories, $\mathcal W$, $\mathcal X$, and $\mathcal Z$. The set $\mathcal W$ contains the histories that The set $\mathcal Z$ contains the histories that could be recorded if we only knew when each individual was truly captured. could be observed during the experiment (the recorded histories). The set $\mathcal X$ contains the histories that could be recorded if we knew when each individual was truly captured and also when errors in identification occurred. To distinguish between the final two sets we call elements of $\mathcal Z$ latent capture histories and elements of $\mathcal X$ latent error histories. As in the example, we use $\bm \omega$, $\bm \nu$, and $\bm \xi$ to identify individual elements of $\mathcal W$, $\mathcal X$, and $\mathcal Z$ respectively. Counts of these histories are denoted by $\bm n$, $\bm x$, and $\bm z$, and we index these vectors both in order and by name. For example, $z_i$ represents the count for the $i^{th}$ latent capture history using some ordering and $z_{\bm \xi}$ represents the count of $\bm \xi \in \mathcal Z$.  

Generally, we let $\bm n$ be the $I$-vector of counts for the observable histories, $\bm x$ the $J$-vector of counts for the latent error histories, and $\bm z$ the $K$-vector of counts for the latent capture histories. As in \citet{Link2010}, we assume that $\bm n=\bm A \bm x$ for some known matrix $\bm A$. Further, we assume that $\bm z=\bm B \bm x$ for some known matrix $\bm B$. The complete data likelihood is then constructed in two stages: 1) modelling the process of capturing, marking, and recapturing individuals to define $\pi(\bm z|\bm \theta)$ and 2) modelling the error process conditional on the true captures to define $\pi(\bm x|\bm z,\bm \theta)$. We expect that the parameters in these components will be separate so that we can represent them by the disjoint sets $\bm \theta_1$ and $\bm \theta_2$. 
% The complete data likelihood is then defined by the product
% \[
% \pi(\bm n,\bm x,\bm z|\bm \theta_1,\bm \theta_2)=\ind(\bm n=\bm A \bm x)\pi(\bm x|\bm z,\bm \theta_2)\pi(\bm z|\bm \theta_1)
% \]
% and the joint posterior distribution is
The posterior distribution of the complete data and parameters is 
\[
\pi(\bm x,\bm z,\bm \theta_1,\bm \theta_2|\bm n) \propto \ind(\bm n=\bm A \bm x)\pi(\bm x|\bm z,\bm \theta_2)\pi(\bm z|\bm \theta_1)\pi(\bm \theta_1)\pi(\bm \theta_2)
\]
where $\pi(\bm \theta_1)$ and $\pi(\bm \theta_2)$ represent priors assumed to be independent. When considering a specific history (observed or latent) we index the vectors of counts both by index and by name. For example, $n_i$ represents the count for the $i$th observable history, using some implicit ordering, while $n_{\bm \omega}$ represents the count of history $\bm \omega$. A table summarizing the notation is provided in Appendix \ref{app:notation}.

To fit the \CJSBRE into the extended framework we need to  1) identify the sets of observable histories, latent error histories, and latent capture histories; 2) construct the constraint matrices; and 3) define the components of the likelihood function. 
% To simplify notation we introduce the entry $\cdot$ into the capture histories to identify the events up to and including the time of marking which do not contribute to the likelihood. For example, the latent capture history $\bm \xi_i=\cdot\cdot101$ indicates that individual $i$ was marked on occasion 2 and recaptured on occasions 3 and 5 of a study with $T=5$ occasions. 
Conditioning on the first release we can exclude both the unmarked individuals and the individuals marked on the final occasion from the likelihood. This leaves $I=2^T-2$ observable histories containing the events 0 and 1 excluding the zero history and the history ending with a single capture. The latent capture histories also belong to the same set so that $K=2^T-2$ as well. In defining the latent error histories, four events can occur on each occasion after marking. The $i$th individual may be not resighted (event 0), resighted and correctly identified (event 1) or resighted and incorrectly identified (event 2). Another marked individual may be captured and incorrectly identified as individual $i$ (event 3). Events 2 and 3 represent false negative and false positive resightings. 

\paragraph{Constraint Matrices: A}
Next, we construct the constraint matrices $\bm A $ and $\bm B$. One factor that makes the \CJSBRE model more complicated than model \Mta{} is that it contains constraints on $\bm x$ beyond those imposed by the observed counts. In particular, $\pi(\bm x|\bm z,\bm \theta_2)>0$ only if the number of false positives and false negative captures are equal on all occasions. The $\bm A$ matrix is constructed as
\[
\bm A=
\begin{bmatrix}
  \bm A_1\\
  \bm A_2
\end{bmatrix}
\]
where $\bm A_1$ is a $(2^T-2) \times J$ matrix modelling the relationship between $\bm x$ and $\bm n$ that is defined similar to the matrix $\bm A'$ in \citet[][]{Link2010}, and $\bm A_2$ is a $(T-1) \times J$ matrix constraining the number of false positives and negatives on the final $T-1$ occasions. Mathematically,
\[
A_{1ij}=
\left\{
  \begin{array}{ll}
    1 & \mbox{if } \omega_{it}=\ind(\nu_{jt}=1) + \ind(\nu_{jt}=3) \mbox{ for all } t=1,\ldots,T\\
    %& \mbox{or if } \bm \omega_{i}=\bm e_t \mbox{ and } \bm \nu_{j}=2\bm e_t \mbox{ for some } t=1,\ldots,T\\
    0 & \mbox{otherwise}
  \end{array}
\right.
\]
% so that $A_{1ij}=1$ if $\bm \omega_i$ is observed from $\bm \nu_j$ and $A_{1ij}=0$ otherwise. Then
and
\[
A_{2tj}=\left\{
  \begin{array}{ll}
    -1 & \mbox{if } \nu_{j,t+1}=2\\
    1 & \mbox{if } \nu_{j,t+1}=3\\
    0 & \mbox{otherwise}
  \end{array}
\right.
\]
The $t$th row of $\bm A_2$ computes the difference between the number of 2s and 3s in the latent error histories, and the vector $\bm n$ must also be extended by concatenating $T-1$ extra 0s corresponding to the added constraints. The $\bm B$ matrix is defined such that $B_{jk}=1$ if the $j$th latent capture history has the same pattern of captures as the $k$th latent error history. That is
\[
B_{jk}=
  \left\{
    \begin{array}{ll}
      1 & \mbox{if } \xi_{kt}=\ind(\nu_{jt}=1) + \ind(\nu_{jt}=2) \mbox{ for all } t=1,\ldots,T\\
      0  & \mbox{otherwise}
    \end{array}
  \right..
\]

\paragraph{Likelihood Function}
Finally, we define the distributions of $\bm z$ and $\bm x|\bm z$. Let $a_t$ denote the number of individuals first captured and marked on occasion $t$, $M_t$ the number of individuals marked before occasion $t$, and $m_t$ the number of these individuals resighted on occasion $t$. The density of $\bm z$ is a product multinomial
  \[
  \pi(\bm z|\bm \theta_1)=
  \frac{\prod_{t=1}^{T-1} a_{t}!}{\prod_{k=1}^K \bm z_k!} \prod_{k=1}^K pr(\bm \xi_k|\bm \theta_1)^{z_k}
  \]
where $pr(\bm \xi_k|\bm \theta_1)$ denotes the probability assigned to history $\bm \xi_k$ by the \CJS model. To construct the second component of the likelihood we consider occasions $t=2,\ldots,T$ separately first modelling the number of errors that occur, denoted by $E_t$, and then modelling the exact configuration of false positives and false negatives. The assumptions of the \BRE model imply that $E_t \leq m^*_t=\min(m_t,M_t-m_t)$ and so we model $E_t$ according to the (possibly) truncated binomial with density
\[
pr(E_t=e_t|\bm z,\alpha) \propto
{m_t \choose e_t} (1-\alpha)^{e_t}\alpha^{m_t-e_t}
% {\sum_{e=0}^{\min(m_t,M_t-m_t)} {m_t \choose e} (1-\alpha)^{e_t}\alpha^{m_t-e}}
, \quad e_t=0,\ldots,m^*_t.
\]
where $\alpha$ is the probability of a correct identification. We assume that all assignments of false positives and false negatives are equally likely conditional on $E_t$. There are ${m_t \choose E_t}$ and ${M_t-m_t \choose E_t}$ ways to select the false negatives and false positives and it follows that 
\[
pr(\bm x|E_2,\ldots,E_T,\bm z)=\ind(\bm z= \bm B \bm x) \prod_{t=2}^T\left[{m_t \choose E_t}{M_t-m_t \choose E_t}\right]^{-1}.
\]
The second component of the likelihood is
\[
\pi(\bm x|\bm z,\alpha)= \ind(\bm z= \bm B \bm x)
\frac{\prod_{k=1}^K \bm z_{k}!}{\prod_{j=1}^J \bm x_{j}!}
\prod_{t=2}^T \left[
  \frac{(1-\alpha)^{e_t(\bm x)}\alpha^{m_t-e_t(\bm x)}}
  {{M_t-m_t \choose e_t(\bm x)}\sum_{e=0}^{m^*_t} {m_t \choose e} (1-\alpha)^{e_t(\bm x)}\alpha^{m_t-e}}
\right]
\]
% \[
% \pi(\bm x|\bm z,\alpha)= \ind(\bm z= \bm B \bm x)
% \frac{\prod_{k=1}^K \bm z_{k}!}{\prod_{j=1}^J \bm x_{j}!}
% \prod_{t=2}^T \left[
%   \frac{(1-\alpha)^{e_t(\bm x)}\alpha^{m_t-e_t(\bm x)}}
%   {\sum_{e=0}^{\min(m_t,M_t-m_t)} {m_t \choose e} (1-\alpha)^{e_t(\bm x)}\alpha^{m_t-e}}
%   {M_t-m_t \choose e_t(\bm x)}^{-1}
% \right]
% \]
where $e_t(\bm x)=\sum_{j=1}^J x_{j} \ind(\nu_{jt}=2)$ represents the number of errors in configuration $\bm x$. The initial term accounts for the many relabellings of the marked individuals that would produce the same counts in $\bm x$ and $\bm z$.

\section{Dynamic Markov Bases}

\label{sec:methods-2}

% \subsection{Dynamic Markov Bases}
% \label{sec:dmb}

% Furthermore, many of the moves within a particular Markov subbasis will lead to proposals with negative entries when combined with the current counts of the latent histories, $\bm x^{(k-1)}$, regardless of the coefficient, $c$. Suppose that on some particular iteration of the algorithm both $x^{(k-1)}_{j}=0$ and $x^{(k-1)}_{k}=0$. Any move, $\bm b$, such that $b_{j}=-1$ and $b_{k}=1$ will lead to a proposal with $x^{\mathrm{prop}}_{j}<0$ if $c>0$ or $x^{\mathrm{prop}}_{k}<0$ if $c<0$. This move can be  ignored -- though it may be used on later iterations.

As in \citet{Link2010}, the main difficulty in sampling from the joint posterior distribution of $\bm \theta_1$, $\bm \theta_2$, $\bm x$, and $\bm z$ lies in generating proposals for $\bm x$ that are inside $\mathcal F_{\bm n}$ with high probability. The addition of the second vector of latent counts does not complicate matters because $\bm z$ is a deterministic function of $\bm x$. This means that consistent proposals for $\bm x$ and $\bm z$ can be constructed jointly by generating a proposal for $\bm x$ and then setting $\bm z=\bm B \bm x$. 

We initially tried to construct Markov bases with the software package \texttt{4ti2} which uses general methods based on the theory of toric ideals to compute Markov bases. Unfortunately, the Markov bases for the \CJSBRE model grow so quickly with $T$ that they cannot be computed for studies of a reasonable size.  For $T \geq 5$, \texttt{4ti2} ran out of memory on a computer with 8~GB of RAM before completing the calculations. 

To avoid this problem we make use of dynamic Markov bases. \citet{Dobra2012} defined a dynamic Markov bases to be a collection of sets of local moves, $\mathcal M(\bm x)$, which connect each $\bm x \in \mathcal F_{\bm n}$ to a relatively small number of neighbours. On the $k$th iteration of the \MCMC algoirthm a proposal is generated by sampling a move from $\mathcal (\bm x^{(k-1)}$). This avoids the need to compute the entire Markov basis and increases the efficiency of the algorithm because the local moves are more likely to generate proposals inside the fibre. 

\citet{Dobra2003} initially proposed a method for dynamically generating moves for the problem of sampling from a table of arbitrary dimension with constraints on some marginal sums. In particular, they showed that the set of primitive moves containing two 1s and two -1s forms a Markov basis for such models and developed an algorithm for dynamically selecting moves from this basis. \citet{Dobra2012} generalized this to the problem of sampling tables of arbitrary dimension constrained by fixing some marginal sums and also placing bounds on individual cell entries. The \CJSBRE models does not fit into the framework because the constraints on the number of false negatives and false positives do not correspond to marginal sums or bounds on individual cells. However, a dynamic Markov basis containing only primitive moves can still be constructed as follows.

Define $\mathcal X_{vt}(\bm x)=\left\{\bm \nu:\nu_t=v \mbox{ and } x_{\bm \nu} >0 \right\}$ to be the subset of latent error histories with event $v$ on occasion $t$ and positive entries in $\bm x$. The moves in $\mathcal M(\bm x)$, denoted by $\bm b(\bm \nu_0,\bm \nu_1,\bm \nu_2,\bm \nu_3)$, each contain two -1s associated with histories $\bm \nu_0 \in \mathcal X_{0t}(\bm x)$ and $\bm \nu_1 \in \mathcal X_{1t}(\bm x)$ and two 1s associated with histories $\bm \nu_2 \in \mathcal X_{2t}(\bm x)$ and $\bm \nu_3 \in \mathcal X_{3t}(\bm x)$ for some common $t$. The moves can be defined by drawing pairs of elements from any two of the four sets, but it makes most sense to consider drawing the histories from $\mathcal X_{0t}(\bm x)$ and $\mathcal X_{1t}(\bm x)$ and then constructing the histories in $\mathcal X_{2t}(\bm x)$ and $\mathcal X_{3t}(\bm x)$ or vice versa. Given histories $\bm \nu_0 \in \mathcal X_{0t}(\bm x)$ and $\bm \nu_1 \in \mathcal X_{1t}(\bm x)$ the histories $\bm \nu_2 \in \mathcal X_{2t}(\bm x)$ and $\bm \nu_3 \in \mathcal X_{3t}(\bm x)$ are constructed by setting
\[
\nu_{2t}=\left\{
  \begin{array}{cl}
    \nu_{0s} & \mbox{if } s \neq t\\
    2 & \mbox{if } s=t
  \end{array}
  \right.
\mbox{ and }
\nu_{3t}=\left\{
  \begin{array}{cl}
    \nu_{3t} & \mbox{if } s \neq t\\
    3 & \mbox{if } s=t
  \end{array}
  \right..
\]
More compactly, $\bm \nu_2=\bm \nu_0 + 2 \bm \delta_t$ and $\bm \nu_3=\bm \nu_2 + 2 \bm \delta_t$ where $\bm \delta_t$ represents the $J$-vector with a single 1 in entry $t$. The entry of $\bm b(\bm \nu_0,\bm \nu_1,\bm \nu_2,\bm \nu_3)$ associated with the latent error history $\bm \nu$ is
\begin{equation}
\label{eq:bre-move}
b_{\bm \nu}(\bm \nu_0,\bm \nu_1,\bm \nu_2,\bm \nu_3)=
\left\{
  \begin{array}{rl}
    -1 & \mbox{if } \bm \nu=\bm \nu_0 \mbox{ or } \bm \nu=\bm \nu_1\\
    1 & \mbox{if } \bm \nu=\bm \nu_2 \mbox{ or } \bm \nu=\bm \nu_3\\
    0 & \mbox{if } \mathrm{Otherwise}
  \end{array}
\right..
\end{equation}
Alternatively, the same set of moves can be obtained by pairing all $\bm \nu_2 \in \mathcal X_{2t}(\bm x)$ and $\bm \nu_3 \in \mathcal X_{3t}(\bm x)$, setting $\bm \nu_0=\bm \nu_2 - 2 \bm \delta_t$ and $\bm \nu_1=\bm \nu_3 - 2 \bm \delta_t$, and defining the entries $\bm b(\bm \nu_0,\bm \nu_1,\bm \nu_2,\bm \nu_3)$ exactly as above. Heuristically, the operation $\bm x + c~\bm b(\bm \nu_0,\bm \nu_1,\bm \nu_2,\bm \nu_3)$ adds $c$ errors when $c>0$ and $c$ removes errors when $c<0$.

For each $\bm x \in \mathcal F_{\bm n}$ , $\mathcal M(\bm x)$ is the union of two sets of moves constructed as in equation (\ref{eq:bre-move}):
%\begin{align*}
\[
\mathcal{M}_1(\bm x)=\{\bm b(\bm \nu_0,\bm \nu_1,\bm \nu_2,\bm \nu_3):\bm \nu_0 \in \mathcal X_{0t}(\bm x) \mbox{ and } \bm \nu_1 \in \mathcal X_{1t}(\bm x) \mbox{ for some } t\}
\]
and
\[
\mathcal{M}_2(\bm x)=\{\bm b(\bm \nu_0,\bm \nu_1,\bm \nu_2,\bm \nu_3):\bm \nu_2 \in \mathcal X_{2t}(\bm x) \mbox{ and } \bm \nu_3 \in \mathcal X_{3t}(\bm x) \mbox{ for some } t\}.
\]
The first set contains all moves that incorporate new errors when added to $\bm x$ while keeping the proposal inside $\mathcal F_{\bm n}$. The second contains all moves that remove errors when subtracted from $\bm x$ and again keep the proposal inside $\mathcal F_{\bm n}$. On the $k$th iteration a proposal for $\bm x$ is then generated by sampling $c \in \mathcal \{-1,1\}$, selecting 
\[
\bm b \in \left\{
  \begin{array}{ll}
    \mathcal M_1(\bm x^{(k-1)}) & \mbox{if } c=-1\\ 
    \mathcal M_2(\bm x^{(k-1)}) & \mbox{if } c=1
  \end{array}
\right.,
\]
and setting $\bm x^{\mathrm{prop}}=\bm x^{(k-1)} + c\bm b$.
% \setcounter{step}{0}
% \begin{step}
%   sampling $c \in \mathcal \{-1,1\}$,
% \end{step}
% \begin{step}
%   selecting 
%   $\bm b \in \left\{
%   \begin{array}{ll}
%     \mathcal M_1(\bm x^{(k-1)}) & \mbox{if } c=-1\\ 
%     \mathcal M_2(\bm x^{(k-1)}) & \mbox{if } c=1
%     \end{array}
%   \right.$, and
% \end{step}
% \begin{step}
%   setting $\bm x^{\mathrm{prop}}=\bm x^{(k-1)} + c\bm b$.
% \end{step}
% \begin{enumerate}
% \item sampling $c \in \mathcal \{-1,1\}$,
% \item selecting 
% $\bm b \in \left\{
%   \begin{array}{ll}
%     \mathcal M_1(\bm x^{(k-1)}) & c>0\\ 
%     \mathcal M_2(\bm x^{(k-1)}) & c<0
%     \end{array}
%   \right.$, and
% \item setting $\bm x^{\mathrm{prop}}=\bm x^{(k-1)} + c\bm b$.
% \end{enumerate}
\noindent
Full details are provided in Algorithm \ref{alg:3}. That the algorithm connects all elements in $\mathcal F_{\bm n}$ is given by Theorem \ref{thm:3}. 
\begin{theorem}
  \label{thm:3}
  Let $\bm x_1,\bm x_2 \in \mathcal F_{\bm n}$. Then there exists a sequence of moves $\bm b_1,\ldots,\bm b_L$ and coefficients $c_1,\ldots,c_L\in \{-1,1\}$, for some $L$, such that $\bm x_2=\bm x_1 + \sum_{l=1}^L c_l \bm b_l$, $\bm x_1 + \sum_{l=1}^{L'} c_l \bm b_l \in \mathcal F_{\bm n}$, and
  \[\bm b_l \in \mathcal M\left(\bm x_1 + \sum_{l=1}^{L'} \bm b_l \right)\]
for all $L'=1,\ldots,L$.
\end{theorem}
\noindent
The proof in Appendix \ref{app:proof} shows that any element in $\mathcal F_{\bm n}$ can be connected to the unique configuration with no errors, denoted by $\bm x^{\bm 0}$, by subtracting elements from $\mathcal M_2(\bm x)$ to remove errors one at a time. Any other element in $\mathcal F_{\bm n}$ can then be reached by adding elements from $\mathcal M_1(\bm x)$ to add errors one at a time.

One further advantage of this approach is that moves can be sampled from $\mathcal M(\bm x)$ without ever having to compute this set explicitly. If $c=1$ then $\bm b \in \mathcal M_1(\bm x)$ can be drawn by: 1) selecting $\bm \nu_0 \in \bigcup_{t=1}^T \mathcal{X}_{0t}(\bm x)$, 2)sampling $s \in \{t:\nu_{0t}=2\}$, and 3) selecting $\bm \nu_1\in \mathcal X_{1s}(\bm x)$. If $c=-1$ then $\bm b \in \mathcal M_1(\bm x)$ can be drawn by: 1) selecting $\bm \nu_2 \in \bigcup_{t=1}^T \mathcal{X}_{2t}(\bm x)$, 2) sampling $s \in \{t: \nu_{2t}=0\}$, and 3) selecting $\bm \nu_3 \in \mathcal X_{3s}(\bm x)$. 
% \setcounter{step}{0}
% \begin{step}
%  selecting $\bm \nu_0 \in \bigcup_{t=1}^T \mathcal{X}_{0t}(\bm x)$,
% \label{step1}
% \end{step}
% \begin{step}
%  sampling $s \in \{t:\nu_{0t}=2\}$, and
% \end{step}
% \begin{step}
%  selecting $\bm \nu_1\in \mathcal X_{1s}(\bm x)$.
% \label{step3}
% \end{step}
% \noindent
% If $c=-1$ then a move can be chosen by
% \setcounter{step}{0}
% \begin{step}
%  selecting $\bm \nu_2 \in \bigcup_{t=1}^T \mathcal{X}_{2t}(\bm x)$,
% \end{step}
% \begin{step}
%  sampling $s \in \{t: \nu_{2t}=0\}$, and
% \end{step}
% \begin{step}
%  selecting $\bm \nu_3 \in \mathcal X_{3s}(\bm x)$.
% \end{step}
In either case the proposal density, $q(\bm x'|\bm x)$, is proportional to the inverse of the product of the cardinalities of the sets in each of the three steps. It is possible that the one of three sets may be empty. This happens if we try to add an error when no correct identifications are available (if $c>0$) or to remove an error when no errors exist ($c<0$). In these cases we retain $\bm x^{(k-1)}$ with probability one and continue to the next iteration. 

\begin{algorithm}
  \caption{Proposed algorithm for sampling from the posterior distribution of $\bm \theta_1$, $\bm \theta_2$, $\bm x$, and $\bm z$ using dynamic moves.}
  Initialize $\bm \theta_1^{(0)}$, $\bm \theta_2^{(0)}$, $\bm x^{(0)}$, and $\bm z^{(0)}$ so that $\bm n = \bm A \bm x^{(0)}$ and $\bm z^{(0)}=\bm B \bm x^{(0)}.$\;
 
\BlankLine 
Set $k=1$.\;

%\BlankLine
  
%\Repeat{Convergence}{
%\Indp
  1) Update $\bm \theta_1$ and $\bm \theta_2$ conditional on $\bm x^{(k-1)}$ and $\bm z^{(k-1)}$. Call the results $\bm \theta_1^{(k)}$ and $\bm \theta_2^{(k)}$.\;
  \BlankLine
  2) Update $\bm x$ conditional on $\bm \theta_1^{(k)}$ and $\bm \theta_2^{(k)}$ as follows.\;
  \Indp
  %\Indp
  a) Sample $c \in \{-1,1\}$ and $\bm b \in \mathcal M(\bm x^{(k-1)})$.\;
  b) Set $\bm x^{\mathrm{prop}}= \bm x^{(k-1)} + c\bm b$ and ${\bm z}^{\mathrm{prop}}=\bm B {\bm x}^{\mathrm{prop}}$.\;
  c) Calculate the Metropolis acceptance probability: 
  \[
  r = \min\left\{1,\frac{\pi(\bm n,\bm x^{\mathrm{prop}},{\bm z}^{\mathrm{prop}}|\bm \theta_1^{(k)},\bm \theta_2^{(k)})}{\pi(\bm n,\bm x^{(k-1)},{\bm z}^{(k-1)}|\bm \theta_1^{(k)},\bm \theta_2^{(k))})}\cdot 
    \frac{q(\bm x^{(k-1)}|\bm x^{\mathrm{prop}})}{q(\bm x^{\mathrm{prop}}|\bm x^{(k-1)})}
  \right\}
  \]
  where $q(\bm x'|\bm x)$ is the probability of proposing $\bm x'$ given the current state $\bm x$.\;
  d) Set $\bm x^{(k)}=\bm x^{\mathrm{prop}}$ and $\bm z^{(k)}=\bm z^{\mathrm{prop}}$ with probability $r$. Otherwise, set $\bm x^{(k)}=\bm x^{(k-1)}$ and $\bm z^{(k)}=\bm z^{(k-1)}$.\;
  \BlankLine
  
  %\Indm
  \Indm
  3) Increment $k$.
  \BlankLine
%}
%%% Local Variables: 
%%% mode: latex
%%% TeX-master: "../bre_manuscript_1"
%%% End: 

  \label{alg:3}
\end{algorithm}

% \subsection{Markov Bases}

% \label{sec:markov-bases}

Although we were not able to compute a minimal Markov bases for the \CJSBRE model  with the software package \texttt{4ti2} we can construct a full Markov basis as the set of all moves that add or remove errors. Samples from the joint posterior distribution of $\bm \theta$, $\bm x$, and $\bm z$ for the \CJSBRE model could then, in principle, be generated using this Markov basis in Algorithm \ref{alg:1} with an additional step to compute $\bm z^{\mathrm{prop}}=\bm B \bm x^{\mathrm{prop}}$ as in Algorithm \ref{alg:3}. However, this Markov basis is so large that the approach is entirely impractical. It contains more than $6.41 \times 10^{10}$ elements when $T=10$. A matrix of this size would require 128~GB of memory even when stored as two lists of indices identifying the 1s and -1s, and it is highly unlikely (almost impossible) that a randomly selected move would produce a proposal inside $\mathcal F_{\bm n}$. 

\section{Computational Efficiency}

\label{sec:simulation-1}

To illustrate the gain in efficiency from using our dynamic Markov basis we present results from analyzing a single simulated data set with $T=4$ capture occasions (the largest number for which we can compute the Markov basis using \texttt{4ti2}). Data was generated for a sample of 30 individuals, 10 released on each of the first three capture occasions, with constant survival probability $\phi_1=\phi_2=\phi_3=.8$, constant capture probability $p_2=p_3=p_4=.5$, and error rate $\alpha=.5$. Samples from the joint posterior distribution of $\bm x$ and $\bm z$ were then drawn using both the Algorithm \ref{alg:1} which uses the fixed Markov bases, modified for the extended framework as described in Section \ref{sec:methods-1}, and Algorithm \ref{alg:3} which generates moves dynamically. The parameters, $\bm \theta$, were fixed at their true values.

% {\it One issue in comparing the algorithms directly is that Algorithm \ref{alg:2} cycles through all 1683 elements in the Markov basis on each iteration whereas Algorithm \ref{alg:3} generates only one proposal for $\bm x$ per iteration. To account for this, we measured the length of the chains in terms of the number of proposals considered and not the number of iterations. Both chains were run until 10,000 proposals had been generated and either accepted or rejected. In the case of Algorithm \ref{alg:2} we retained the sample obtained on each accept/reject step instead of keeping only the values generated at the end of each iterations (i.e., every 1683 accept/reject steps).}

\begin{figure}
  %\centering
  
  \input{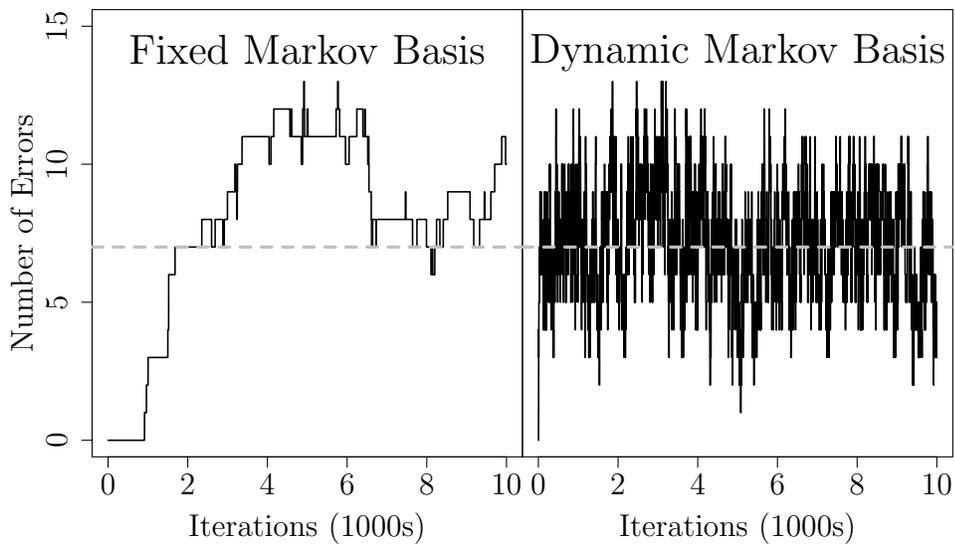}

  \caption{Comparison of the chains sampling from the posterior distribution of the \BRE model applied to the simulated data. The figures trace the number of errors in $\bm x$ for the algorithms using the fixed basis (left) and dynamic basis (right). The grey dotted lines represent the true number of errors in the data set.}
  \label{fig:ex2-1}
\end{figure}

To assess how well the two chains mixed we compared the acceptance rates and the number of unique solutions for $\bm x$ identified per accept/reject step. The chain constructed using Algorithm \ref{alg:1} identified a total of 79 unique configurations among the 7,500 values of $\bm x$ sampled after the burn-in phase. Less than $1\%$ of the proposed configurations were accepted. In comparison, the chain constructed with Algorithm \ref{alg:3} identified 2548 unique configurations and $38\%$ of the proposed configurations were accepted. Figure \ref{fig:ex2-1} also provides traceplots of the number of errors in the configurations sampled by the two chains on each accept/reject step. These summaries all make it clear that the chain constructed from Algorithm \ref{alg:3} is mixing and moving through the fibre much more quickly than the chain constructed from Algorithm 1 \ref{alg:1}.

%%% mode: latex
%%% TeX-master: "../bre_manuscript_1"
%%% End:

% \section{Example 1: \Mta}

% \input{examples_1}

% \section{Example 2: Band-Read-Errors}

% \input{examples_2}

\section{Results}

\label{sec:application}

In our analysis of the queen snake data we fit an initial \CJS model to the original PIT tag data (Model 1). We then simulated data mimicking what might be observed from the branding data by adding errors in the PIT tag data according to the BRE model using the observed identification rates, $\alpha=8/9$ and $\alpha=6/9$, and refit the \CJS model to each data set to assess the effects of errors that are not modeled (Model 2). Finally, we fit the BRE model to the corrupted data sets using our \MCMC implementation with dynamic moves (Model 3). The data for all models included 10 capture occasions: the marking session in 2013, the marking session in spring of 2014, and the 8 visits to resight/recapture snakes in the summer of 2014. One hundred simulated data sets were generated for each value of $\alpha$. 

\begin{figure}
  \centering

  \input{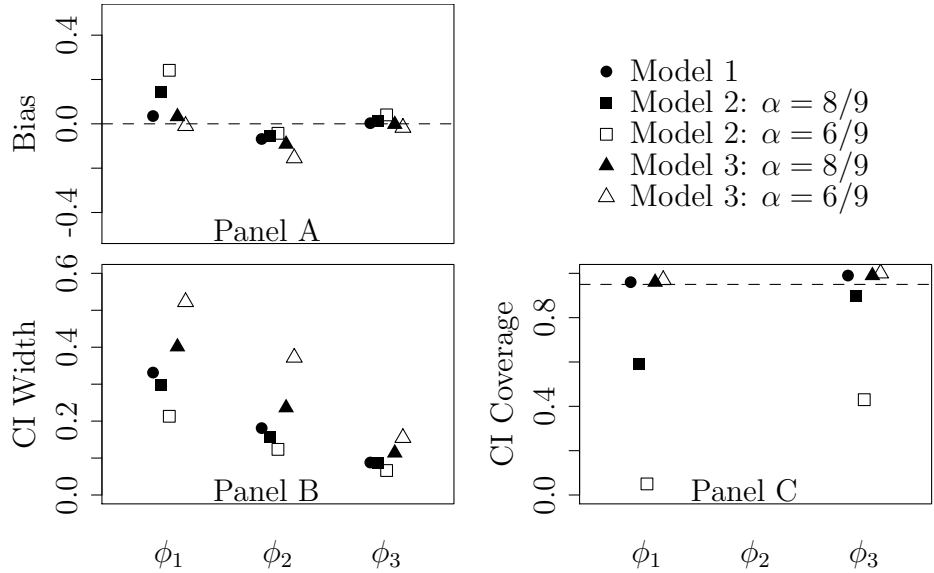}
  % \figurebox{.01in}{6in}{}[dummy.eps]

  \caption{Results of the analysis of the queen snake data. The three panels present the estimate bias of the posterior means (Panel A) and the estimated width (Panel B) and coverage probability (Panel C) of the 95\% credible interval for the survival probabilities for the three models described in Section \ref{sec:application}. The different models are indicated by the shape of the plotting symbol. The rates of error are for Models 2 and 3 indicated by the color of the symbol. Coverage of $\phi_2$ is not reported because the true parameter lies on the boundary of the parameter space.}
  \label{fig:sim-1}
\end{figure}

Parameters for the data generating models were based on analysis of the original PIT tagging data. Analysis of the PIT tag data using classical, maximum likelihood methods in Program MARK \citep{White1999} strongly supported a simplified \CJS model which allowed the capture probabilities to vary independently across all occasions but constrained survival to be equal across the final 8 occasions. Specifically, the estimated survival probabilities were $\hat \phi_1=0.66$, $\hat \phi_2=1.00$, and $\phi_3=\ldots=\phi_9=0.93$. The first of these represents the overwintering survival, the probability that a snake marked in 2013 is still alive in the study area in the summer of 2014, and is of primary interest. The capture and survival probabilities were assigned uniform prior distributions in all models. When fitting Model 2 and Model 3 we assumed that $\alpha$ was known. The correct identification rate would need to be estimated in practice but the data are likely contain little information about this parameter, as noted by \citet{Link2010}. To avoid problems with identifiability \citet{Link2010} assign $\alpha$ an informative  $\beta(19,1)$ prior distribution which places 95\% of its mass between 0.85 and 1.00. Alternatively, one could use double observers, repeated identifications, or double marks to estimate $\alpha$ directly.  

Figure \ref{fig:sim-1} summarizes the posterior distributions for the survival probabilities obtained from the three models. In particular, we compare the bias of the posterior means and the width and coverage of the central 95\% credible intervals. The coverage of $\phi_2$ is zero for all models and is not reported because the parameter lies on the boundary of the parameter space. 

The most important differences are for the estimation of the overwintering survival rate, $\phi_1$. Fitting the data without errors (Model 1) provides estimates of all three parameters that are almost unbiased. Coverage probabilities of $\phi_1$ and $\phi_3$ are also above the nominal value. Ignoring the errors and fitting the same \CJS model (Model 2) produced very poor results. When $\alpha=8/9$ the estimated bias in the overwintering survival rate was 0.15 (23\%) and the coverage of this parameter was only 47\%. The bias increased one and a half times to 0.24 (36\%) when the correct identification probability decreased to $\alpha=6/9$ and the coverage dropped to only 6\%. Although posterior means of $\phi_1$ and $\phi_3$ from Model 2 were not significantly biased, the credible intervals were too narrow and the coverage of $\phi_3$ was still very low when $\alpha=8/9$. In comparison, point estimates of $\phi_1$ and $\phi_3$ from the \CJSBRE model (Model 3) were negligibly biased for both levels of error. Coverage of these parameters was almost identical for Models 1 and 3 and exceeded the nominal rate. All models produced biased estimates of $\phi_2$, which is not surprising given that the true parameter lies on the boundary of the parameter space. The posterior mean of $\phi_2$ from Model 3 was significantly more biased than that of Model 1, underestimating $\phi_2$ by 9\% when $\alpha=8/9$ and 16\% when $\alpha=6/9$. This is due to there being more significant shrinkage toward the prior mean of 0.50 when there is more uncertainty in the data. As expected, credible intervals from Model 3 wider than those from Model 1 to account for the extra uncertainty introduced by the errors. 

% % latex table generated in R 3.1.2 by xtable 1.7-1 package
% % Mon Nov 17 14:00:46 2014
% \begin{table}
% \caption{Results of the analysis of the queen snake data. The columns in the table present the estimate bias of the posterior means and the estimated width and coverage probability of the 95\% credible interval for the survival probabilities for the three models described in Section \ref{sec:application}. Note that the coverage of $\phi_2$ is not reported because this parameter lies on the boundary of the parameter space.
% } 

% \centering
% \begin{tabular}{llrrrrrrrrr}
%   \hline
%   & & \multicolumn{3}{c}{Bias}& \multicolumn{3}{c}{CI Width}& \multicolumn{3}{c}{CI Coverage} \\ 
%  & & $ \phi_1$ &$ \phi_2$ &$ \phi_3$ &$ \phi_1$ &$ \phi_2$ &$ \phi_3$ &$ \phi_1$ &$ \phi_2$ &$ \phi_3$ \\ 
%  \hline
% Model 1 &  & 0.04 & -0.07 &  0.00 &  0.33 &  0.19 &  0.09 &  0.96 &  NA &  0.98 \\ 
% \\
%   Model 2: & $\alpha=8/9$ &  0.14 & -0.06 &  0.01 &  0.30 &  0.16 &  0.09 &  0.59 & NA &  0.90 \\
%   & $\alpha=6/9$ &  0.24 & -0.04 &  0.04 &  0.22 &  0.12 &  0.06 &  0.06 &  NA &  0.42 \\ 
%   \\
%   Model 3: & $\alpha=8/9$ &  0.03 & -0.09 &  0.00 &  0.40 &  0.24 &  0.11 &  0.96 &  NA &  0.99 \\  
%   & $\alpha=6/9$ & -0.01 & -0.16 & -0.01 &  0.52 &  0.38 &  0.15 &  0.96 &  NA &  1.00 \\ 
%    \hline
% \end{tabular}
% \label{tab:sim-2}
% \end{table}

%%% Local Variables: 
%%% mode: latex
%%% TeX-master: "../bre_manuscript_1"
%%% End: 

\section{Discussion}

\label{sec:discussion}

The results in Section \ref{sec:application} clearly illustrate the problems with misidentification. The overwintering survival probability was overestimated by 23\% when the error rate was low and 36\% when the error rate was high. In comparison, estimates from the \CJSBRE model were almost unbiased and the credible intervals had above nominal coverage. The extra uncertainty in the error does increase the posterior variances and so the obvious recommendation is to reduce error rates experimentally by using marks that are easier to identify, double tagging individuals, or using multiple observers. Uncertainty could also be reduced pairing observers to decrease the error rates or simply by increasing the number of hours spent in the field to raise capture rates. Of course, these options would require more man power and increase the cost of branding as a method of marking individuals. We are currently assessing these options to find the optimal method of studying the queen snake population.  

We have described our methods specifically for the \CJSBRE model, but we believe that the methods should be applicable to broad range of mark-recapture models with possible identification errors. The extended framework described in Section \ref{sec:methods-1} can incorporate more complex models of both the capture and error processes than the original \LMM and is particularly useful when the distribution of the joint histories described by the combining processes is complicated or intractable. The algorithm based on dynamic Markov bases presented in Section \ref{sec:methods-2} essentially entails moving through the fibre, $\mathcal F_{\bm n}$, by adding or removing errors one at a time, and we expect that the same procedure can be applied widely. There are two important caveats. First, it must be possible to write the model in terms of the two linear constraints described in the extended framework. This will not always be the case and does not happen if we extend the \BRE model so that individual $i$ can be captured on occasion $t$ and another individual can be captured and identified as individual $i$ at the same time. We are working to extend these models to allow for such events. The second caveat is that the Markov chains derived from the new algorithm may not be irreducible if the posterior distribution assigns probability zero to some elements in the fibber. This might occur if certain configurations of the errors can be ruled out {\it a priori}. 

Another issue that remains to be addressed is the level of connectivity and efficiency of the chains produced with the different algorithms. Any Markov basis is guaranteed to connect all elements in $\mathcal F_{\bm n}$ for any $\bm n$, but the chains produced by different algorithms will connect the fibre in different ways and the result chains will mix more or less well. As an extreme example, a certain Markov basis may divide the fibre into two parts each with many internal connections but with only one connection between. In this case the fibre is connected but any Markov chain generated by this basis will tend to remain in one of the two parts and mix poorly. These issues are complex and there is very little literature within the field of algebraic statistics providing metrics to compare the connectivity for different Markov bases or assessing the effects on efficiency. We hope that further developments in this area will provide guidance on choosing between Markov bases that we can apply to the problem of misidentification in mark-recapture studies. 

% \begin{itemize}
% \item Fewster's paper implementing \Mta
%   \begin{itemize}
%   \item Too specific and can't be generalized.
%   \end{itemize}
% \end{itemize}
%%% Local Variables: 
%%% mode: latex
%%% TeX-master: "../bre_manuscript_1"
%%% End: 

\appendix

\section{Notation}
\label{app:notation}

% \begin{table}
%   \caption{Summary of notation.}

\begin{longtable}{p{.5in}p{4.25in}}
  %\label{tab:notation}
  \multicolumn{2}{l}{\underline{Algebraic Statistics and Markov Bases:}}\\
  $\mathcal F_{\bm n}$ & $n$-fiber: $\mathcal F_{\bm n}=\{\bm x \in \mathbb N^J: \bm n=\bm A\bm x\}$.\\
  $\mathcal B$ & Lattice basis for $\ker(\bm A)$.\\
  $\mathcal M$ & Markov basis for $\ker(\bm A)$.\\
  $\mathcal M_{\bm n}$ & Markov subbasis for $\ker(\bm A)$ given $\bm n$.\\
  $\mathcal M(\bm x)$ & Dynamic Markov basis for $\ker(\bm A)$ computed at $\bm x$.\\
  \\
  \\
  \multicolumn{2}{l}{\underline{Extended \LMM:}}\\
  $\bm \omega_i$ & Observed capture history for the $i^{th}$ marked individual. The subscript is dropped when computing probabilities for a general history.\\
  $\bm \nu_i$ & Latent error history for the $i^{th}$ marked individual.\\
  $\bm \xi_i$ & Latent capture history for the $i^{th}$ marked individual.\\
  $\bm n$ & Known vector of counts for the observed histories (indexed by $i$ and $\bm \omega$).\\
  $\bm x$ & Unknown vector of counts for the latent error histories (indexed by $j$ and $\bm \nu$).\\
  $\bm z$ & Unknown vector of counts for the latent capture histories (indexed by $k$ and $\bm \xi$).\\
  $I$ & Length of $\bm n$. For the \CJSBRE model $I=2^T-2$.\\
  $J$ & Length of $\bm x$. For the \CJSBRE model $J=(4^T-1)/3-1$.\\
  $K$ & Length of $\bm z$. For the \CJSBRE model $K=2^T-2$.\\
  $\bm A$ & $I \times J$ matrix mapping $\bm x$ onto $\bm n$, $\bm n=\bm A\bm x$.\\
  $\bm B$ & $K \times J$ matrix mapping $\bm x$ onto $\bm z$, $\bm z=\bm B\bm z$.\\
  $\bm \theta_1$ & Parameters in the model of $\bm z$.\\
  $\bm \theta_2$ & Parameters in the conditional model of $\bm x$ given $\bm z$.\\
  
\\
\\
\multicolumn{2}{l}{\underline{Band-Read Error Model:}}\\
$p_t$ & Capture probability: the probability that an individual alive on occasion $t$ is captured, $t=2,\ldots,T$. \\
$\phi_t$ & Survival probability: the probability that an individual is alive on occasion $t+1$ given that it was alive on occasion $t$, $t=1,\ldots,T-1$.\\
$\alpha$ & Correct identification rate: the probability that a captured individual is identified correctly.\\

\end{longtable}
% \end{table}

\section{Proof of Theorem \ref{thm:3}}
\label{app:proof}

The proof that the sets $\mathcal M(\bm x)$ defined in Section \ref{sec:methods-2} connect $\mathcal F_{\bm n}$ follows by (reverse) induction on the number of errors. Let $\bm x \in \mathcal F_{\bm n}$ and suppose that in this configuration the number of errors on occasion $t$ is greater than 0: $\sum_{j=1}^J x^{\bm r}_j \ind(\nu_{jt}=2)=\sum_{j=1}^J x^{\bm r}_j \ind(\nu_{jt}=3)>0$. Then there exist histories $\bm \nu_2 \in \mathcal X_{2t}(\bm x)$ and $\bm \nu_3 \in \mathcal X_{3t}(\bm x)$ such that $x_{\bm \nu_2}>0$, $x_{\bm \nu_3}>0$. We can subtract the move $\bm b(\bm \nu_0,\bm \nu_1,\bm \nu_2,\bm \nu_3)$ to reach a new element $\mathcal F_{\bm n}$ with one less error on occasion $t$. Repeating this procedure, we reach an element in $\mathcal F_{\bm n}$ with no errors. To complete the proof, we note that the only element in $\mathcal F_{\bm n}$ with no errors is $\bm x^{\bm 0}$ where
\[
x^{\bm 0}_{\bm \nu}=
\left\{
  \begin{array}{ll}
    n_{\bm \nu} & \mbox{if $\bm \nu$ is observable}\\
    0 & \mbox{otherwise}.
  \end{array}
\right.
\]
It follows that every history can be connected to $\bm x^0$ using moves from the dynamic Markov basis and that the partial sums remain inside the fiber. Following the path in the reverse order by adding elements from $\mathcal M_1(\bm x)$ we can reach any other element in the fiber. 

%%% Local Variables: 
%%% mode: latex
%%% TeX-master: "../bre_manuscript_1"
%%% End: 

\bibliographystyle{SJBbiometrics}
\bibliography{../bre_manuscript_1}

\end{document}